\documentclass[aps,pra,showpacs,twocolumn,floats]{revtex4}
\usepackage{amssymb}
\usepackage{amsbsy}
\usepackage{amsmath}
\usepackage{graphicx}
\usepackage{amsmath}
\usepackage{times}
\usepackage{color}
\usepackage{setspace}
\usepackage{bm}
\newcommand\bea{\begin{eqnarray}}
\newcommand\eea{\end{eqnarray}}
\newcommand\beq{\begin{equation}}
\newcommand\eeq{\end{equation}}

\begin{document}

\title{{\em Zitterbewegung} in Spin-Orbit Coupled Systems and Ehrenfest's Theorem}

\author{Debabrata Sinha and Jayanta K. Bhattacharjee}

\affiliation{Theoretical Physics Department, Indian Association for
the Cultivation of Science, Jadavpur, Kolkata-700032, India.}

\date{\today}

\begin{abstract}
We use Ehrenfest's theorem to provide a particularly simple derivation of the {\em zitterbewegung} in the dynamics of initial Gaussian wave packets in a two dimensional electron gas. For initial packets which are very wide in the $y$-direction, the {\em zitterbewegung} is only in the $y$-component of the velocity. We extend our Ehrenfest theorem based calculation to the spin-orbit coupled spinor Bose-Einstein condensate (BEC) to predict that there can be {\em zitterbewegung} in the $x$-component of the velocity in this situation driven by a combination of the nonlinear interaction in the condensate and the splitting due to the spin-orbit coupling.

\end{abstract}
\maketitle
\section{Introduction}
Ehrenfest's theorem in quantum mechanics, in its generalized form, gives the dynamics of the average value $\langle \mathcal{O}\rangle$ of an arbitrary operator $\mathcal{O}$ evolving under the action of a Hamiltonian $H$. For an operator with no explicit time dependence, one has,
\begin{eqnarray}
i\hbar \frac{d\langle \mathcal{O}\rangle}{dt}=\langle [\mathcal{O},H] \rangle
\label{Ehren}
\label{Ehrenfest}
\end{eqnarray}
If $\mathcal{O}$ is the position operator $\vec{r}$, the dynamics leads to $\frac{d}{dt}\langle \vec{p}\rangle=0$ for the free particle and to $\frac{d}{dt}\langle \vec{p}\rangle=-m\omega^2 \langle \vec{r} \rangle$ for a free particle moving in a simple harmonic potential. Since these equations for the average values resemble the classical equations of motion, Ehrenfest's theorem has often been linked to the classical limit. For a one dimensional motion in the potential $V(x)=\frac{\lambda}{4}x^4$, it is apparent, that the Ehrenfest dynamics for $\langle p\rangle$ does not corresponds to the classical dynamics. However, introducing a further constraint that the fluctuation around the mean is small, one can make Ehrenfest's theorem look approximately like the classical limit \cite{Shankar}. It has been made amply clear in Ref\cite{Wheeler}. that Ehrenfest's theorem contains a huge amount of information about the state of quantum particle. Our contention here is that Eq.(\ref{Ehrenfest}) has to be taken at its face value and if the dynamics of a given moment connects to a different (often higher) moment, then one should look at the dynamics of the new moment to eventually arrive at a closed system after a few attempts or failing which set up a closure scheme to obtain the dynamics of the different moments. Consequently, looking at Ehrenfest's equations can lead to a different perspective on quantum dynamics where the dynamics of at least a few low order moments may be followed quite accurately. Generally moments are what can be easily usually studied in an experiment, and consequently this should be a useful viewpoint. Since nothing can be more quantum mechanical than {\em Zitterbewegung}\cite{Zawaclzki} and in the last decade there has been a resurgence of interest in this unexpected phenomenon, we focus in this work on using Ehrenfest's theorem to deal with {\em Zitterbewegung}.

{\em Zitterbewegung}\cite{Zawaclzki} (trembling motion) was first predicted by Schr\"{o}dinger\cite{Schr} as a consequence of the interference between the particle and hole components of the spinors describing the wave functions of relativistic Dirac electrons in the absence of any external potential. It has never been observed in this setting. However in the last two decades different approaches, originating in non-relativistic quantum mechanics, have been developed to get analytic, numerical and experimental handels on the phenomenon of {\em Zitterbewegung}. One of early suggestions in this direction was to exploit the intrinsic spin orbit interaction existing in low dimensional systems. Electron wave packet dynamics including the issue of {\em Zitterbewegung} in semiconductor quantum well with Rashba and Dresselhaus coupling was studied by Schliemann {\em et.al.}\cite{Schliemann-PRL,Schliemann-PRB} in the absence of electric and magnetic fields. A different study, considering the interplay between spin orbit coupling and cyclotron motion in a perpendicular magnetic field was carried out by Winkler {\em et.al.}\cite{Winkler-PRB,Ziilicke}. Demikhovskii {\em et.al.}\cite{Frolova-PRB08} have carried out a detailed analytic and numerical investigation of the wave packet dynamics of one and two-dimensional wave packets in a semiconductor quantum well under the influence of the Rashba spin orbit coupling. The splitting of an initial wave packet and the accompanying {\em Zitterbewegung} had a particularly elegant expressions in the special situation where the initial wave packet in the two dimensional plane had a very large width in the $y$-direction compared to its width in the $x$-direction. These results were later extended to study the dynamics of wavepackets in a monolayer graphene\cite{Frolova-PRB}. A simulation of the Dirac equation using a single trapped ion was carried out by Gerritsma {\em et.al.}\cite{Gerritsma} to show the existence of {\em Zitterbewegung} in a situation which carefully mimics the Dirac equation.

For systems with Rashba spin orbit coupling, the Hamiltonian studied by Demikhovskii {\em et.al.}\cite{Frolova-PRB08} can be written as,
\begin{eqnarray}
H=H_0+H_R=\frac{p^2}{2m}+\alpha(p_y\sigma_x-p_x\sigma_y)
\label{spin-orbit-hamil}
\end{eqnarray}
where $H_0=\frac{p^2}{2m}$ is the usual free particle Hamiltonian while the remaining $H_R$ is the Rashba spin-orbit term. This Hamiltonian has eigenvalues
\begin{eqnarray}
E(p)=\frac{p^2}{2m}\pm \alpha |p|
\end{eqnarray}
with the eigenvalues $\Psi_1$, $\Psi_2$ respectively,
\begin{eqnarray*}
\Psi_1&=&e^{i\vec{p}\cdot \vec{r}}\begin{pmatrix} 1\\ -ie^{i\phi}\end{pmatrix}\\
\Psi_2&=&e^{i\vec{p}\cdot \vec{r}}\begin{pmatrix} 1\\ ie^{i\phi}\end{pmatrix}
\end{eqnarray*}
where $\phi$ is the angle between $\vec{p}$ and the $x$-axis.  

Demikhovskii {\em et.al.}\cite{Frolova-PRB08} consider the initial Gaussian wave-packet
\begin{eqnarray} 
\Psi_{0}=\frac{1}{\sqrt{\pi d \Delta_0}}e^{-\frac{x^2}{2d^2}}e^{-\frac{y^2}{2\Delta^2_0}}e^{ik_0x}\begin{pmatrix} 1\\ 0 \end{pmatrix}
\label{psi-zero}
\end{eqnarray}
and in the special (and rather illuminating) situation of $\Delta_0>>d$ (no $y$-dependence) effectively obtain explicitly the wave function at later time (using the usual Green function approach)
\begin{eqnarray}
\Psi(x,t)=\begin{pmatrix} \psi_1(x,t) \\ \psi_2(x,t) \end{pmatrix}=\begin{pmatrix} (f(x,t)+g(x,t)e^{i\phi(x,t)})\\ (f(x,t)-g(x,t)e^{i\phi(x,t)})\end{pmatrix}
\label{two-comp-wave}
\end{eqnarray}
where 
\begin{eqnarray}
f(x,t)&=&\frac{C}{\sqrt{\Delta(t)}}e^{-[x+(\alpha-\frac{\hbar k_0}{m})t]^2/2\Delta^2}\nonumber\\
g(x,t)&=&\frac{C}{\sqrt{\Delta(t)}}e^{-[x-(\alpha+\frac{\hbar k_0}{m})t]^2/2\Delta^2}\nonumber\\
\phi(x,t)&=&\frac{2(k_0d^2+\frac{\hbar t}{md^2}x)\alpha t}{\Delta^2}\nonumber\\
\label{Delta-Square}
\Delta^2(t)&=&d^2+\frac{\hbar^2 t^2}{m^2d^2}
\end{eqnarray}
In the above $C$ is a normalization constant. The forms of $f(x,t)$ and $g(x,t)$ which are centred at $(\frac{\hbar k_0}{m}\pm \alpha)t$ show the splitting of the wave packet. The spin density is
\begin{eqnarray*}
S_i(x,t)=\frac{\hbar}{2}(\Psi^*_1 \Psi^*_2)\sigma_i \begin{pmatrix} \Psi_1\\\Psi_2\end{pmatrix}
\end{eqnarray*}
The spin density $S_y(x,t)$ is seen to be
\begin{eqnarray}
S_y(x,t)&=&\frac{\hbar}{\sqrt{\pi}L_y \Delta}[e^{-[x-(v_0-\alpha)t]^2/\Delta^2}\nonumber\\&&-e^{-[x-(v_0+\alpha)t]^2/\Delta^2}]
\end{eqnarray}
where $v_0=\frac{\hbar k_0}{m}$ and the average spin density 
\begin{eqnarray}
\bar{S}_y=\int S_y(x,t) dx dy =0
\label{avg-sy}
\end{eqnarray}
The average spin density $\bar{S}_z$ oscillate in time. Further, it was shown that
\begin{eqnarray}
\langle p_x \rangle&=&\frac{\hbar k_0}{m}\\
\label{avg-v}
\langle v_y \rangle&=&-\alpha \sin(2k_0\alpha t)e^{-(\frac{\alpha t}{d})^2}
\end{eqnarray}
The oscillating but decaying $y$-velocity is the {\em Zitterbewegung}. Our objective in Sec.II will be to arrive at the above results from the prospective of Ehrenfest's theorem in a very straightforward manner.

The actual experimental observation \cite{Qu,Blanc} of {\em Zitterbewegung} was achieved by considering spin-orbit coupled Bose-Einstein condensates\cite{Zhai-PRL10}. The dynamics of the condensate, described by the two component wave function $\begin{pmatrix} \Psi_1\\ \Psi_2 \end{pmatrix}$ is described by the coupled Gross-Pitaevskii equation which can be written as,
\begin{eqnarray}
i\hbar {\partial \over \partial t}\begin{pmatrix} \psi_1\\ \psi_2 \end{pmatrix}&=&[\frac{p^2}{2m}+\alpha(p_y\sigma_x-p_x\sigma_y)\begin{pmatrix} \psi_1\\ \psi_2 \end{pmatrix}\nonumber\\
&&+ G\begin{pmatrix} \psi_1\\ \psi_2 \end{pmatrix}]
\label{spin-BE}
\end{eqnarray}
with 
\begin{eqnarray}
G=\begin{pmatrix} g_{11}|\psi_1|^2+g_{12}|\psi_2|^2 & 0\\ 0 & g_{22}|\psi_2|^2+g_{12}|\psi_1|^2 \end{pmatrix}
\label{int-term}
\end{eqnarray}
In both references Ref.\cite{Qu}. and Ref.\cite{Blanc}, a clear oscillation of the average velocity is observed in the absence of any external forces. In Ref.\cite{Qu}, the observations actually exploited the spin-orbit coupling and it is this situation which we will concentrate on. The use of Ehrenfest's equation will show that the mean position of the wave packet will primarily be the same as that shown in Eq.(\ref{two-comp-wave}), although the width of the packet is capable of showing deviation from that shown in Eq.(\ref{Delta-Square}). It will be seen that unless the interaction between the atoms of the condensate is attractive, the effect on the width will be qualitatively unaltered. The most spectacular effect of the interaction between atoms, we predict, will be a possible non zero value of the average spin $\bar{S}_y$ if the coefficients $g_{\alpha \beta}$ of Eq.(\ref{int-term}) satisfy certain constraints. This will be the content of our Sec.III. We conclude with a short summary in Sec.IV.

\section{Ehrenfest equations for the electron gas}
Following the geometry studied by Demikhovskii {\em et.al.}\cite{Frolova-PRB08}, we consider the Hamiltonian of Eq.(\ref{spin-orbit-hamil}) and the initial condition of Eq.(\ref{psi-zero}) with $\Delta_0>> d$ so that the time development wave packet has only a $x$-dependence. Consequently, we will have $\langle p_y \rangle =0$ all the time- a fact that will be used throughout in writing down expectation values. We can immediately write down the following results using Ehrenfest's theorem
\begin{eqnarray}
\label{vx-eq}
\langle v_x \rangle&=&\frac{d}{dt}\langle x \rangle=\frac{\langle p_x \rangle}{m}-\alpha \langle \sigma_y \rangle\\
\label{vy-eq}
\langle v_y \rangle&=&\frac{d}{dt}\langle y \rangle=\alpha \langle \sigma_x \rangle\\
\label{p-eq}
\frac{d}{dt}\langle p_x \rangle&=&\frac{d}{dt}\langle p_y \rangle=0
\end{eqnarray}
In this case $\langle p_y \rangle=0$, and if initially $\langle p_x \rangle=p_0=\hbar k_0/m$, then $\langle p_x \rangle$ remains at $p_0$ all through. The important thing to note is that although $\langle p_y \rangle=0$, $\langle v_y \rangle \neq 0$ and this will give one of the primary results, exhibited in Eq.(\ref{avg-v}). Continuing with the application of Eq.(\ref{Ehren}), one get
\begin{eqnarray}
\frac{d}{dt}\langle \sigma_x \rangle&=&-2\alpha \langle p_x\sigma_z \rangle\nonumber\\
&=&-2\alpha[\int \psi^*_1 p_x\psi_1 dx-\int \psi^*_2 p_x\psi_2 dx]\nonumber\\
&=&-2\alpha[\langle p_x \rangle_{11}-\langle p_x \rangle_{22}]
\label{sigmax-dyn}
\end{eqnarray}
where the meaning of $\langle p_x \rangle_{11}$ and $\langle p_x \rangle_{22}$ is obvious. Similarly,
\begin{eqnarray}
\frac{d}{dt}\langle \sigma_y \rangle=-2\alpha \langle p_y \sigma_z \rangle=0
\label{sigmay-time}
\end{eqnarray}
since both $\langle p_y \rangle_{11}$ and $\langle p_y \rangle_{22}$ vanish, the functions $\Psi_1$ and $\Psi_2$ being functions of $x$ alone. Hence $\langle \sigma_y \rangle$ is constant in time and can be evaluted from its value at $t=0$. The initial wavefunction is given in Eq.(\ref{psi-zero}) with $\Delta_0 \rightarrow \infty$ and hence has the form
\begin{eqnarray}
\Psi_{0}(x)=\Psi (x,t=0)=\frac{1}{\pi^{\frac{1}{4}}\sqrt{d}}e^{-\frac{x^2}{2d^2}}\begin{pmatrix} 1\\ 0\end{pmatrix}
\label{psi-cor}
\end{eqnarray}
Since $\langle \sigma_y \rangle (t=0)=-i\int \psi^*_1\psi_2 dx+i \int \psi^*_2 \psi_1 dx$, $\langle \sigma_y \rangle (t=0)=0$ and thus
\begin{eqnarray*}
\langle \sigma_y \rangle =0
\end{eqnarray*}
for all time. Immediately it follows that
\begin{eqnarray}
\langle v_x \rangle =\frac{p_0}{m}=\frac{\hbar k_0}{m}
\label{avg-vx-eq}
\end{eqnarray}
Thus we have, thanks to Ehrenfest's theorem, obtained two of the results of Ref\cite{Frolova-PRB08}. as given in Eq.(\ref{avg-sy}) and Eq.(\ref{avg-v}) without any serious calculations.

We now turn to Eq.(\ref{sigmax-dyn}). The right hand side is trivially calculated in momentum space. The momentum space wave function at $t=0$ (corresponding to Eq.(\ref{psi-cor})) is
\begin{eqnarray}
\Phi_{0}(k)=\begin{pmatrix} \phi_1(k)\\ \phi_2(k) \end{pmatrix}=\sqrt{2}d \pi^{\frac{1}{4}}e^{-(k-k_0)^2d^2/2}\begin{pmatrix} 1\\ 0 \end{pmatrix}
\end{eqnarray}
The initial case is a superposition of $\begin{pmatrix} 1 \\ -i\end{pmatrix}$ and $\begin{pmatrix} 1 \\ i\end{pmatrix}$, the eigenstates with eigenvalues $\frac{\hbar^2k^2}{2m}+\alpha \hbar k$ and $\frac{\hbar^2k^2}{2m}-\alpha \hbar k$, respectively it follows that at time t, the momentum space wave function is
\begin{eqnarray}
\Phi(k)=\sqrt{2}d\pi^{\frac{1}{4}}e^{-(k-k_0)^2d^2/2}e^{i\frac{\hbar k^2 t}{2m}}\begin{pmatrix} \cos (\alpha k t) \\ \sin(\alpha k t)
\end{pmatrix}
\label{mom-wave}
\end{eqnarray}
Now evaluating $\langle p_x \rangle_{11}$ and $\langle p_x \rangle_{22}$, we find from Eq.(\ref{vy-eq}) and Eq.(\ref{sigmax-dyn})
\begin{eqnarray}
\frac{d}{dt}\langle v_y \rangle=\alpha e^{-\frac{\alpha^2t^2}{d^2}}[k_0d \cos(2\alpha k_0 t)-\frac{\alpha t}{d}\sin(2\alpha k_0 t)]
\end{eqnarray}
leading on integration to
\begin{eqnarray}
\langle v_y \rangle=-\alpha \sin(2\alpha k_0 t) e^{-\frac{\alpha^2 t^2}{d^2}}
\end{eqnarray}
the oscillating but decaying $y$-component of velocity shown in Eq.(\ref{avg-v}). This is the appearance of {\em Zitterbewegung} in the two dimensional electron gas with spin orbit coupling.

One can actually make further progress within the Ehrenfest framework. For the width of the wave packet one has on evaluating the relevant commutators
\begin{eqnarray}
\frac{d}{dt}\langle x^2 \rangle=\frac{1}{m}\langle xp_x+p_x x\rangle-2\alpha\langle x\sigma_y \rangle
\label{xsq-eq}
\end{eqnarray}
and
\begin{eqnarray}
\frac{d}{dt}\langle xp_x+p_x x\rangle=\frac{2}{m}\langle p^2_x\rangle-2\alpha\langle \sigma_y p_x \rangle
\label{xp-eq}
\end{eqnarray}
Clearly, $\frac{d}{dt}\langle p^2_x\rangle=0$ and for wave functions of the form $f(x)\begin{pmatrix}\alpha \\ \beta\end{pmatrix}$, $\frac{d}{dt}\langle \sigma_y p_{x}\rangle=0$, making the right hand side of Eq.(\ref{xp-eq}) constant. Thus taking a derivative of Eq.(\ref{xsq-eq}) and using Eq.(\ref{xp-eq}) we get
\begin{eqnarray}
\frac{d^2}{dt^2}\langle x^2 \rangle&=&2\frac{\langle p^2_x\rangle}{m^2}-\frac{4\alpha}{m}\langle \sigma_y p_x \rangle\nonumber\\
&&=2\frac{\langle p^2_x\rangle_0}{m^2}-\frac{4\alpha}{m}\langle \sigma_y p_x \rangle_0
\label{scxsq-eq}
\end{eqnarray}
where the subscript denote the values at $t=0$. Evaluating the expectation value at $t=0$, one has $\langle p^2_x \rangle_0=\frac{\hbar^2}{2d^2}+\hbar^2k^2_0$ and $\langle \sigma_y p_x \rangle=0$, which gives on integration Eq.(\ref{scxsq-eq})
\begin{eqnarray}
\langle x^2 \rangle=\big(\frac{\hbar^2}{2m^2d^2}+\frac{k^2_0}{m^2}\big)t^2+C_1 t+\frac{d^2}{2}
\end{eqnarray}
where $C_1$ is a constant of motion to be obtained from $\frac{d}{dt}\langle x^2 \rangle$at $t=0$ and is found to be zero. We note (from Eq.(\ref{avg-vx-eq})) that
\begin{eqnarray}
\langle x \rangle=\frac{\hbar k_0 t}{m}
\end{eqnarray}
and hence 
\begin{eqnarray}
\langle x^2 \rangle -\langle x \rangle^2=\frac{\hbar^2}{2m^2d^2}t^2+\frac{d^2}{2}
\end{eqnarray}
If the width of a Gaussian wave packet at any instant is $\Delta(t)$, then $\langle x^2 \rangle -\langle x \rangle^2=\Delta^2/2$ and hence
\begin{eqnarray}
\Delta^2=d^2+\frac{\hbar^2 t^2}{m^2 d^2}
\label{delta-sq}
\end{eqnarray}
which is identical to the width of the wave function obtained in Ref.\cite{Frolova-PRB08} by studying the evolution of the wave packet. 

We would like to end this section by pointing out that although it is not possible to obtain information  on the phase $\phi(x,t)$ of the evolving wave function by Ehrenfest's theorem, in the case of Gaussian wave packets and a free particle an exception may be made. The phase $\phi(x,t)$, if expanded in powers of $'x'$, will not have any powers higher than $x^2$ and we can write
\begin{eqnarray}
\phi(x,t)=\phi_{0}(t)+x\phi_1(t)+x^2\phi_2(x,t)
\end{eqnarray}
In the above, the most important information is carried by the linear term and below we show how information on $\phi_{1}(t)$ can be obtained from a study of the Ehrenfest relation
\begin{eqnarray}
\frac{d}{dt}\langle \sigma_z \rangle=2\alpha \langle p_x\sigma_x \rangle
\end{eqnarray}
The right hand side can be evaluated in momentum space as done before and we find
\begin{eqnarray}
\langle p_x\sigma_x \rangle =e^{-\frac{\alpha^2 t^2}{d^2}}[k_0 d \sin(2\alpha k_0 t)+\frac{\alpha t}{d}\cos (2\alpha k_0 t)]
\end{eqnarray}
The left hand side needs to be evaluated in coordinate space to include the information on phase. We note that because momentum is conserved, the centre of the wave packet is located  at any time $t$ at $x_c=v_0 t=\frac{\hbar k_0 t}{m}$ (as would be true for a free particle without the spin orbit term), but the dispersion relation shows there are two branches, one moving with velocity $v_0+\alpha$ and other with $v_0-\alpha$. Hence the initial Gaussian wave packet will split into two pockets one centred at $(v_0+\alpha)t$ and the other at $(v_0-\alpha)t$ which can be seen by visualizing the Fourier transform of Eq.(\ref{mom-wave}). The real space wave function at any time t, will be a linear superposition of Gaussian centred at $(v_0+\alpha)t$ and $(v_0-\alpha)t$ and having the width $\Delta(t)$ found in Eq.(\ref{delta-sq}) and differing in phase by $\phi(x,t)$. The two component wavefunction in Eq.(\ref{two-comp-wave})
\begin{eqnarray}
\Psi=\begin{pmatrix} \Psi_1(x,t)\\ \Psi_2(x,t)\end{pmatrix}=\begin{pmatrix} f(x,t)+g(x,t) e^{i\phi(x,t)}\\f(x,t)-g(x,t) e^{i\phi(x,t)} \end{pmatrix}
\label{psi-two-compent}
\end{eqnarray}
where
\begin{eqnarray}
\label{f-comp}
f(x,t)&=&C e^{-[x-(v_0+\alpha)t]^2/2\Delta^2}\\
\label{g-comp}
g(x,t)&=&C e^{-[x-(v_0-\alpha)t]^2/2\Delta^2}\\
\phi(x,t)&=& \phi_0(t)+x\phi_1(t)+....
\end{eqnarray}
with $C$ a numerical constant freed by normalization. We find after carrying out the intrigation
\begin{eqnarray}
\frac{d}{dt}\langle \sigma_z \rangle=4\sqrt{\pi}\Delta(t)e^{-\frac{\phi^2_1}{\Delta^2}-\frac{\alpha^2 t^2}{\Delta^2}}[A \cos \phi'+B \sin \phi']
\label{sigmaz-t}
\end{eqnarray}
with 
\begin{eqnarray*}
\phi'&=&\frac{\phi_0}{\Delta^2}+\frac{ip_0t\phi_1}{m\Delta^2}\\
A&=&-\frac{d}{dt}(\frac{\phi^2_1}{\Delta^2}+\frac{\alpha^2 t^2}{\Delta^2})
\end{eqnarray*}
and
\begin{eqnarray*}
B=\frac{d}{dt}\phi'
\end{eqnarray*}
Comparing two sides of Eq.(\ref{sigmaz-t}) leads to 
\begin{eqnarray*}
\phi_0&=&2k_0 d^2\alpha t\\
\phi_1&=&\frac{2\alpha\hbar t^2}{md^2}
\end{eqnarray*}
in agreement with the exact answer of Demikhovskii {\em et.al.}\cite{Frolova-PRB08}.

\section{Spin-Orbit coupled BEC}
We now consider the spin-orbit coupled Bose-Einstein condensates\cite{Zhai-PRL10} of Eq.(\ref{spin-BE}) and once again consider initial wave packets which are the form given in Eq.(\ref{psi-zero}), with $\Delta_0$ made very big so that the dynamics of the packet can be taken to be essentially one dimensional. The extension of Ehrenfest relation to the case of the Gross-Pitaevskii equation was done in Ref.\cite{JKB-IJMP15} and including the spin-orbit coupling, we first write Eq.(\ref{spin-BE}) explicitly as ($g_{11}=g_{22}=g$),
\begin{eqnarray*}
i\hbar \dot{\psi}_1&=&-\frac{\hbar^2}{2m}\nabla^2\psi_1+\alpha (p_y+ip_x)\psi_2\nonumber\\&&+g|\psi_1|^2\psi_1+g_{12}|\psi_2|^2\psi_2\\
i\hbar \dot{\psi}_2&=&-\frac{\hbar^2}{2m}\nabla^2\psi_2+\alpha (p_y-ip_x)\psi_1\nonumber\\&&+g|\psi_2|^2\psi_2+g_{12}|\psi_1|^2\psi_1
\end{eqnarray*}
For operators $\mathcal{O}$ whose expectation values are given by $\langle \mathcal{O} \rangle=\int \Psi^*_1 \mathcal{O}\Psi_1 dx + \int \Psi^*_2 \mathcal{O}\Psi_2 dx$, it was shown \cite{JKB-IJMP15}that the Ehrenfest theorem leads to 
\begin{eqnarray}
i\hbar \langle \dot{\mathcal{O}} \rangle&=&\langle [\mathcal{O},H]\rangle +g[\langle [\mathcal{O},P_1] \rangle_{11}+\mathcal{O},P_2] \rangle_{22}]\nonumber\\&&+g_{12}[\langle [\mathcal{O},P_2] \rangle_{11}+\langle[\mathcal{O},P_1] \rangle_{22}]
\label{mod-Ehren}
\end{eqnarray}
where $H=\frac{p^2}{2m}+\alpha(\sigma_xp_y-\sigma_yp_x)$. It is clear that this $H$ with the spin-orbit coupling does not introduce any new terms in the dynamics of $\langle x \rangle$, $\langle y \rangle$, $\langle p_x \rangle$ and $\langle p_y \rangle$ and Eqs.(\ref{vx-eq},\ref{vy-eq},\ref{p-eq}) continue to hold. The interesting terms arises in the dynamics of $\langle \sigma_x \rangle$ and $\langle \sigma_y \rangle$. We begin with the dynamics of $\langle \sigma_y \rangle$ since without the interaction terms, we had seen that in the situation where $\Psi(x)$ was a function of $x$ alone, $\langle \sigma_y \rangle$ did not change with time. In the present situation, we find that Eq.(\ref{sigmay-time}) changes to,
\begin{eqnarray}
\frac{d}{dt}\langle \sigma_y \rangle=-2\alpha \langle p_y\sigma_z\rangle+(g_{12}-g)\frac{d}{dt}\langle \sigma_y \rangle_{int}
\label{mod-sigma_y}
\end{eqnarray}

where,
\begin{eqnarray}
\frac{d}{dt}\langle \sigma_y \rangle_{int}&=&\int (\psi^*_1\psi_2+\psi^*_2\psi_1)\nonumber\\&&\times(|\psi_1|^2-|\psi_2|^2) dx
\label{mod-sigmay-time}
\end{eqnarray}
The first term in Eq.(\ref{mod-sigma_y}) does not contribute if $\Psi$ depends on $x$-alone, the second is non-vanishing. For small value of $g_{12}$ and $g$, we can evaluate it in the limit of $g=g_{12}=0$ using the known $\Psi_{1,2}(x)$. 
Thus for $g_{12}-g\neq 0$, the average spin in $y$-direction is non-zero. This also has the interesting consequence that the velocity in the $x$-direction has an oscillating behavior in time which eventually damps out. As for $\langle v_y \rangle$ which showed a damped oscillation in time in the absence of $g$, the presence of the interaction term leads to an additional response given by,
\begin{eqnarray}
\frac{d}{dt}\langle \sigma_x \rangle=-2\alpha \langle p_x\sigma_z\rangle+(g-g_{12})\frac{d}{dt}\langle \sigma_x \rangle_{int}
\label{mod-sigma_x-eq}
\end{eqnarray}
where
\begin{eqnarray}
\frac{d}{dt}\langle \sigma_x \rangle_{int}&=&-i\int (\psi^*_1\psi_2-\psi^*_2\psi_1)\nonumber\\&&\times(|\psi_1|^2-|\psi_2|^2) dx
\label{mod-sigmax-time}
\end{eqnarray}
The first term has already been evaluated in Eq.(\ref{sigmax-dyn}). For $g\neq g_{12}$ there is an additional contribution (from the second term in Eq.(\ref{mod-sigma_x-eq}). To evaluate the integrals in Eq.(\ref{mod-sigmay-time}) and Eq.(\ref{mod-sigmax-time}), we make the Gaussian assumption and use the form of $\Psi(x,t)$ as given in Eq.(\ref{psi-two-compent}). We note, at this point, that $\Delta^2(t)$ in this case need not be given by Eq.(\ref{Delta-Square}) since there can be an effect of the interaction on the width. We shall return to this issue later. With $\Psi$ having, the form of Eq.(\ref{psi-two-compent}), we obtain
\begin{eqnarray}
\label{sigma-inty}
\frac{d}{dt}\langle \sigma_y \rangle_{int}&=&-2exp[-\frac{3}{2}\frac{\alpha^2t^2}{\Delta^2(t)}-\frac{\phi^2}{8\Delta^2(t)}]\nonumber\\&&\times\sin(\frac{\phi_0}{\Delta^2(t)}+\frac{\phi_1p_0t}{m\Delta^2(t)})\sin(\frac{\phi_1\alpha t}{2\Delta^2(t)})\\
\label{sigma-intx}
\frac{d}{dt}\langle \sigma_x \rangle_{int}&=&exp[-\frac{2\alpha^2t^2}{\Delta^2(t)}-\frac{\phi^2}{2\Delta^2(t)}]\nonumber\\&&\times\sin(\frac{2\phi_0}{\Delta^2(t)}+\frac{2\phi_1p_0t}{m\Delta^2(t)})
\end{eqnarray}

\begin{figure}
\center
\rotatebox{0}{\includegraphics[width=1.65in]{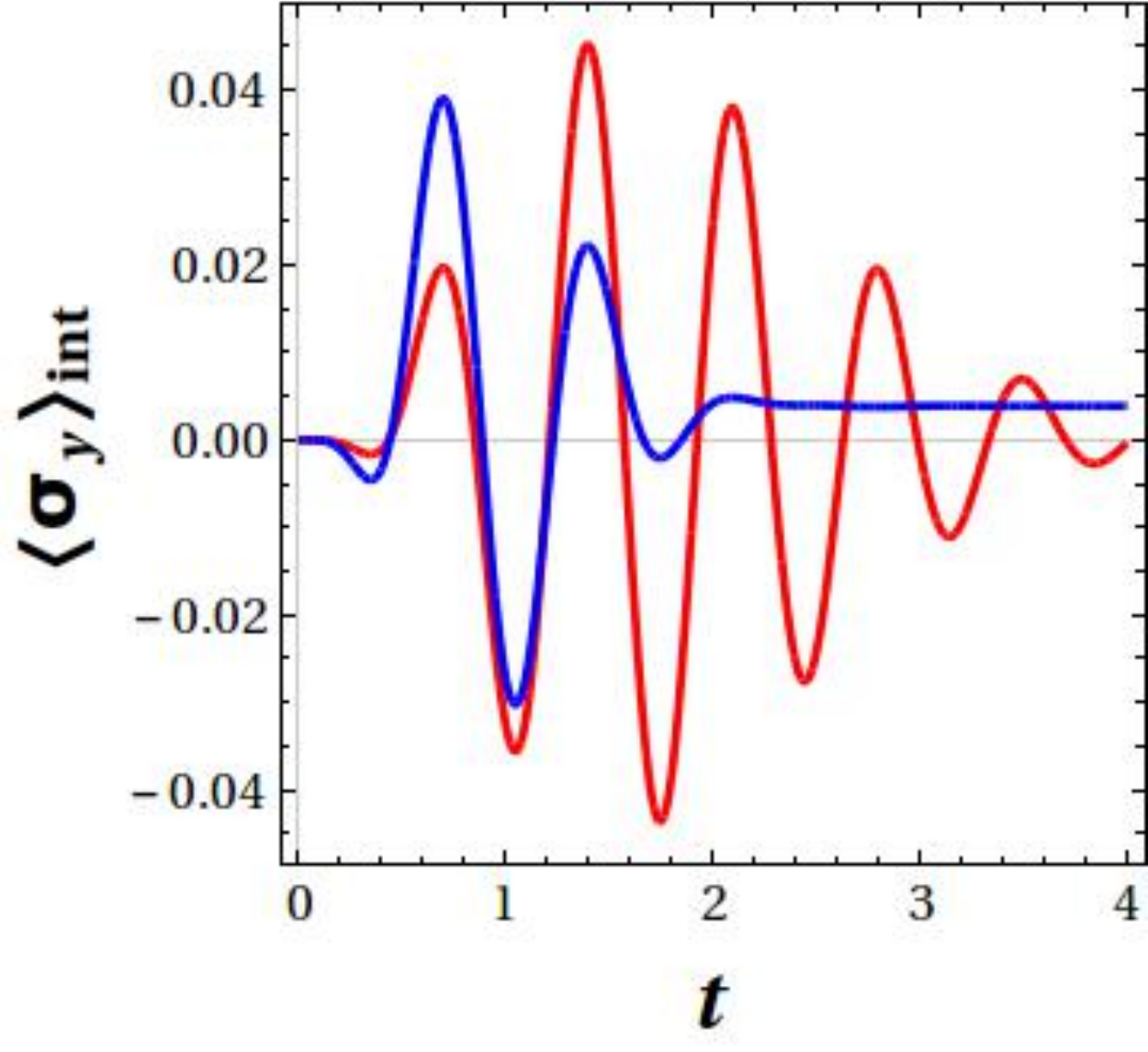}}
\rotatebox{0}{\includegraphics[width=1.6in]{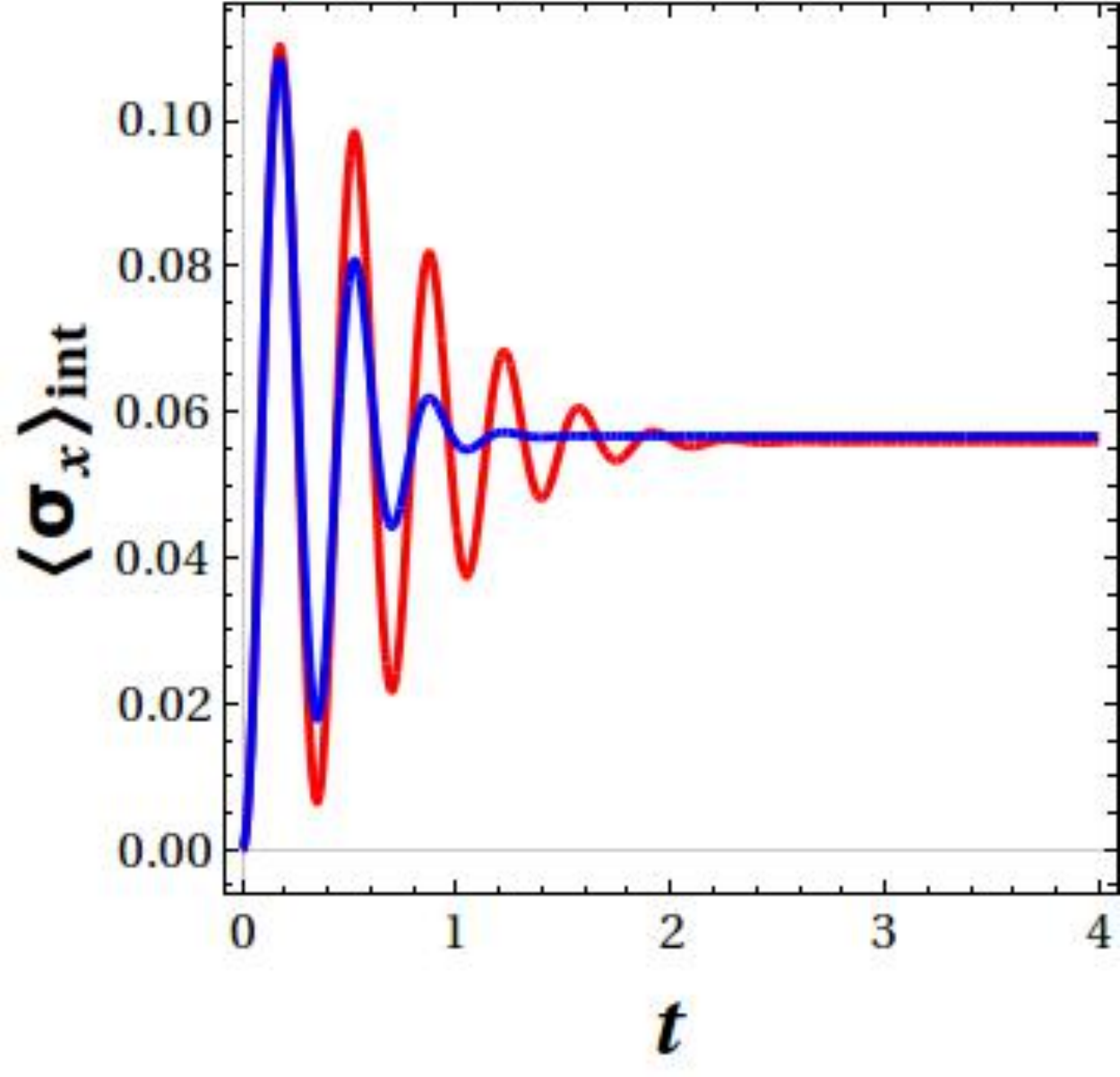}}
\caption{Plot of interaction induced pseudospin components $\langle \sigma_y \rangle_{int}$ and $\langle \sigma_x \rangle_{int}$ with time for different values of $\alpha$ ($\alpha=1.7\times 10^6 cm s^{-1}$ (red) and $\alpha=2.9\times 10^6 cm s^{-1}$(blue) respectively).}
\label{Sigma_comp}
\end{figure}

At this point, we note that most important consequence of the spin-orbit coupling in a BEC. In a situation where the spin-orbit coupling gave a vanishing contribution in the electron gas, namely $\bar{S}_y=0$ for an initial wave-packet which is very wide in the $y$-direction, we find from Eqs.(\ref{mod-sigma_y},\ref{mod-sigma_x-eq}), that the combination of the spin-orbit coupling and the nonlinear interaction can now lead to a non-zero result. Further we also see as a consequence of Eqs.(\ref{mod-sigma_y},\ref{mod-sigma_x-eq}), that the mean velocity $\langle v_x\rangle$ in the $x$-direction will oscillate in time showing the existence of {\em zitterbewegung} in $v_x$ as well in the presence of interaction term. In Fig.(\ref{Sigma_comp}) we plot the components $\langle \sigma_x(t) \rangle_{int}$ and $\langle \sigma_y(t) \rangle_{int}$ determined by Eqs.(\ref{sigma-inty}) and (\ref{sigma-intx}) which demonstrates the effect of {\em zitterbewegung}. We choose other parameters as $m=0.5 Mev/c^2$, $d=10^{-5}cm$ and $k_{0}=2.5\times 10^5cm^{-1}$.

Returning to Eq.(\ref{mod-sigmay-time}), we note that to correctly evaluate the right hand side of Eq.(\ref{mod-sigmay-time}), we need to check whether the interaction have any drastic effect on $\Delta^2(t)$. To this end, we return to Eqs.(\ref{xsq-eq}) and (\ref{xp-eq}) and note that on the inclusion of the interaction terms, Eq.(\ref{xsq-eq}) is unchanged and Eq.(\ref{xp-eq}) becomes
\begin{eqnarray}
\frac{d}{dt}\langle xp_x+p_x x\rangle&=&\frac{2}{m}\langle p^2_x\rangle-2\alpha\langle \sigma_y p_x \rangle\nonumber\\&&+\frac{g}{m}\int (P^2_1+P^2_2) dx +\frac{2g_{12}}{m}\int P_1P_2 dx
\end{eqnarray}
Consequently,
\begin{eqnarray}
\frac{d^2}{dt^2}\langle x^2 \rangle&=&2\frac{\langle p^2_x\rangle}{m^2}-\frac{4\alpha}{m}\langle \sigma_y p_x \rangle\nonumber\\&&+\frac{g}{m}\int (P^2_1+P^2_2) dx +\frac{2g_{12}}{m}\int P_1P_2 dx
\label{xsq-mod-eq}
\end{eqnarray}
For wave function of the form $\frac{d}{dt}\langle \sigma_yp_x\rangle=0$ as before but now $\frac{d}{dt}\langle p^2\rangle$ is no longer zero. With $\langle \sigma_yp_x\rangle$ replaced by its initial value of zero, Eq.(\ref{xsq-mod-eq}) reduces to
\begin{eqnarray}
\frac{d^2}{dt^2}\langle x^2 \rangle&=&2\frac{\langle p^2\rangle}{m^2}+\frac{g}{m}\int (P^2_1+P^2_2) dx\nonumber\\&& +\frac{2g_{12}}{m}\int P_1P_2 dx
\end{eqnarray}
Straightforward algebra leads to
\begin{eqnarray}
\frac{d^2}{dt^2}\langle x \rangle^2=2\frac{\langle p\rangle^2}{m^2}
\end{eqnarray}
The width of the wave packet $W$, defined as $W^2=\langle x^2\rangle-\langle x \rangle^2$, satisfy
\begin{eqnarray}
\frac{d^2}{dt^2}W^2&=&2\frac{\langle (\Delta p\rangle)^2}{m^2}+\frac{g}{m}\int (P^2_1+P^2_2) dx\nonumber\\&&+\frac{2g_{12}}{m}\int P_1P_2 dx
\end{eqnarray}
We need to find the dynamics of $\langle p^2 \rangle$ to be able to analyze the above equation. From Eq.(\ref{mod-Ehren}), we find
\begin{eqnarray}
\frac{d}{dt}\langle p^2\rangle&=&-mg\int {\partial \over \partial t}(P^2_1+P^2_2)dx\nonumber\\&&-2mg_{12}\int {\partial \over \partial t}(P_1P_2)dx
\end{eqnarray}
Since $\frac{d}{dt}\langle p\rangle=0$, we have
\begin{eqnarray}
\frac{d}{dt}\langle (\Delta p)^2\rangle&=&-m\frac{d}{dt}\Big[g\int (P^2_1+P^2_2)dx\nonumber\\&&+2g_{12}\int P_1P_2dx\Big]
\end{eqnarray}
leading to
\begin{eqnarray}
\frac{(\Delta p)^2}{m}+g\int (P^2_1+P^2_2)dx+2g_{12}\int P_1P_2dx=C(const)
\end{eqnarray}
Consequently the dynamics of the width becomes
\begin{eqnarray}
\frac{d^2}{dt^2}W^2=C-g\int (P^2_1+P^2_2)dx-2g_{12}\int P_1P_2dx
\end{eqnarray}
For positive $g$ and $g_{12}$, the constant $C$ is always positive and since $\int (P^2_1+P^2_2)dx$ and $\int P_1P_2dx$ scale as $1/W$, it is clear that as time increases, increasing $W$ is the consistent solution and clearly $W^2\propto Ct^2$ for large $W$. Hence the nature of the width is not expected to change in this case and the conclusions arrive at from Eqs.(\ref{sigma-inty}) and (\ref{sigma-intx}) will be correct.

\section{conclusion}
We have looked at the issue of {\em zitterbewegung} in an electron gas and a spinor BEC in the presence of spin-orbit coupling. The issue of {\em zitterbewegung} is a quintessential quantum phenomenon and we show that in this situation, Ehrenfests theorem (which is often considered as describing the passage to the classical limit) can be very effectively used to arrive at the results obtained by Demikhovskii {\em et. al.}\cite{Frolova-PRB08,Frolova-PRB} for the electron gas. For the more involved case (because of the nonlinear interactions) of the spinor BEC, we find that using Ehrenfest relations one can go beyond the known answers and show that {\em zitterbewegung} can exist in a situation where it did not in the case of electron gas.

\end{document}